\documentclass[twocolumn]{aastex62}
\usepackage{amsmath}     
\usepackage{wasysym}           
\usepackage{graphicx}
\usepackage{amssymb}
\usepackage{epstopdf}
\usepackage{mathrsfs}
\usepackage{anyfontsize}
\usepackage{natbib}
\usepackage{color}
\usepackage{lipsum}
\usepackage{diagbox}

\shorttitle{Partial Stellar Disruption by a Supermassive Black Hole}
\shortauthors{Coughlin \& Nixon}
\begin{document}
\title{Partial Stellar Disruption by a Supermassive Black Hole: Is the Lightcurve Really Proportional to $t^{-9/4}$?}
\author[0000-0003-3765-6401]{Eric R.~Coughlin}
\affiliation{Department of Astrophysical Sciences, Peyton Hall, Princeton University, Princeton, NJ 08544}
\affiliation{Columbia Astrophysics Laboratory, New York, NY 80980}
\author[0000-0002-2137-4146]{C.~J.~Nixon}
\affiliation{Department of Physics and Astronomy, University of Leicester, Leicester, LE1 7RH, UK}

\email{eric.r.coughlin@gmail.com}

\begin{abstract}
The tidal disruption of a star by a supermassive black hole, and the subsequent accretion of the disrupted debris by that black hole, offers a direct means to study the inner regions of otherwise-quiescent galaxies. These tidal disruption events (TDEs) are being discovered at an ever-increasing rate. We present a model for the evolution of the tidally-disrupted debris from a partial TDE, in which a stellar core survives the initial tidal encounter and continues to exert a gravitational influence on the expanding stream of tidally-stripped debris. We use this model to show that the asymptotic fallback rate of material to the black hole in a partial TDE scales as $\propto t^{-2.26\pm0.01}$, and is effectively independent of the mass of the core that survives the encounter; we also estimate the rate at which TDEs approach this asymptotic scaling as a function of the core mass. These findings suggest that the late-time accretion rate onto a black hole from a TDE either declines as $t^{-5/3}$ if the star is completely disrupted or $t^{-9/4}$ if a core is left behind. {We emphasize that previous investigations have not recovered this result due to the assumption of a Keplerian energy-period relationship for the debris orbits, which is no longer valid when a surviving core generates a time-dependent, gravitational potential.} This dichotomy of fallback rates has important implications for the characteristic signatures of TDEs in the current era of wide-field surveys.
\end{abstract}


\section{Introduction}
A star of mass $M_{\star}$ and radius $R_{\star}$ is destroyed by the tidal field of a supermassive black hole (SMBH) of mass $M_{\bullet}$ if the star comes within roughly the tidal radius $r_{\rm t} \equiv R_{\star}\left(M_{\bullet}/M_{\star}\right)^{1/3}$ of the SMBH (e.g., \citealt{hills75, rees88}). Within this distance, the stellar self-gravity is incapable of withstanding the tidal shear of the SMBH, and the star is stretched into a stream of debris \citep{kochanek94, guillochon14, coughlin16b} with roughly half of the debris bound to the SMBH. Upon returning to the point of disruption, the periapsis of the bound stream is advanced through an additional angle owing to general relativistic precession, which causes the material to self-intersect, dissipate energy, and form an accretion flow \citep{hayasaki13, shiokawa15, bonnerot16, hayasaki16, sadowski16}. The accretion episode briefly illuminates the galaxy, and many of these tidal disruption events (TDEs) have now been observed (e.g., \citealt{bloom11, gezari12, cenko12, chornock14, arcavi14, blagorodnova17, hung17, vanvelzen19}; see \citealt{komossa15} for a review).

The characteristic feature of a TDE is often considered to be a smooth decline in luminosity from peak that scales as $t^{-5/3}$, where $t$ is the time since peak. This feature can be derived from the energy-period relationship of a Keplerian orbit (\citealt{phinney89}; see also Section \ref{sec:asymptotic} below), and was reproduced by early simulations of TDEs \citep{evans89}. \citet{lodato09} developed an analytical model based on the impulse, or ``frozen-in,'' approximation, which assumes that the entire star moves with the stellar center of mass at the tidal radius and thereafter the gas parcels comprising the star follow independent Keplerian orbits (see also \citealt{stone13}). They showed analytically that other features of the fallback rate (defined as the rate at which bound, tidally-disrupted material returns to pericenter, which will closely track the true accretion rate if energy is dissipated efficiently), such as the time to peak, depend on the structure of the star (see also \citealt{gallegos-garcia18, steinberg19}). At late times, however, the fallback curves from their analytic model always followed the scaling $\propto t^{-5/3}$, and these analytical predictions were verified with simulations of TDEs. 

The simulations of \citet{lodato09} equated the pericenter distance of the star, $r_{\rm p}$, to the tidal radius, $r_{\rm t}$, and hence the $\beta \equiv r_{\rm t}/r_{\rm p}$ of their encounters -- which measures the degree to which tidal forces overwhelm the stellar self-gravity -- was always equal to one. \citet{guillochon13} performed a suite of hydrodynamical simulations that varied both the stellar structure and $\beta$. They found that there is a critical $\beta$ that separates TDEs into full and partial disruptions, and in the latter scenario a stellar core survives the encounter intact (see also \citealt{mainetti17}). Moreover, \citet{guillochon13} found numerically that the fallback rate from such partial TDEs could be significantly steeper than the expected, $t^{-5/3}$ scaling, and for a range of encounters that left bound cores was better matched by $\sim t^{-2.2}$ (see specifically their Figure 7). 

While \citet{guillochon13} provided estimates for the asymptotic fallback rate from partial TDEs and demonstrated the prolonged influence of the gravitational field of the surviving core, the deviation of the fallback rate from partial TDEs from the canonical, $t^{-5/3}$ scaling has yet to be shown from first principles. Specifically, is it possible to \emph{derive} the asymptotic fallback rate from a partial TDE with a relatively simple physical picture akin to the impulsive model of \citet{lodato09}, and thereby determine the rate as a function of the core mass? In addition, because TDEs are computationally expensive owing to their large temporal and spatial dynamic range, \citet{guillochon13} only simulated the hydrodynamics for the first few days post-disruption (for reference, the most bound material returns to the SMBH on a timescale of $\sim 4$ weeks), and then assumed that the binding energy to the black hole was frozen in to calculate the fallback rate{; those authors also used the energy-period relationship of a Keplerian orbit to forward-predict the late-time fallback, while the presence of a bound core will modify the energy distribution from Keplerian at late times (see below)}. 

In Section \ref{sec:physical} of this Letter, we develop a model for the fallback of tidally-disrupted debris from a partial TDE and show that the asymptotic fallback rate approaches $\propto t^{-2.26\pm 0.01}$, where the upper and lower limits encompass the range of core masses left behind in a typical TDE (see Figure \ref{fig:n_infinity}); we therefore find that partial TDEs are effectively characterized by a universal, late-time fallback rate that scales as $\propto t^{-9/4}$. We also employ a slight variant of the impulse approximation to assess the rate at which partial TDEs conform to this asymptotic limit as a function of the core mass. In Section \ref{sec:summary} we summarize our findings and conclude.

\section{Physical Model}
\label{sec:physical}
We approximate the tidally-disrupted debris stream as a collection of non-self-gravitating fluid elements with a core of mass $M_{\rm c}$ at the zero-energy (with respect to the SMBH) orbit $R(t) \propto t^{2/3}$. We further assume that the stream maintains hydrostatic balance in the non-radial directions, where ``radial'' is in the direction joining the SMBH to the core, and hence the stream is a cylinder of nearly-constant width and each gas parcel possesses a temporally-evolving position $r(t)$ (simulations and analytic arguments have validated these approximations when the stream is adiabatic and gas-pressure dominated; \citealt{guillochon14, coughlin15, coughlin16a, coughlin16b}). The Lagrangian equation of motion for the stream is then\footnote{{Because of the time dependence of the gravitational potential of the core, there is no energy integral of this equation, and the energy distribution of the gas parcels is non-Keplerian; we return to this point in the conclusions.}}

\begin{equation}
\frac{\partial ^2r}{\partial t^2} = -\frac{GM_{\bullet}}{r^2}+\frac{GM_{\rm c}}{\left(R(t)-r\right)^2}\textrm{ sign}\left[R(t)-r\right] , \label{eq1}
\end{equation}
{where here the variable held constant when taking the time derivative is the initial position of the fluid element.} Introducing the change of variables

\begin{equation}
\xi \equiv \frac{r}{R(t)}, \quad \tau \equiv \ln \left(\frac{R(t)}{R_0}\right), \label{xidef}
\end{equation}
where $R_0$ is the initial location of the core, {it follows that}

\begin{equation}
\frac{\partial \xi}{\partial \tau} = \frac{v}{V}-\xi, \label{xioftau}
\end{equation}
{{where $v = \partial r/\partial t$ is the radial velocity and $V(t) = dR/dt$ is the velocity of the core;}} Equation \eqref{eq1} then becomes

\begin{equation}
2\frac{\partial^2\xi}{\partial \tau^2}+\frac{\partial \xi}{\partial \tau}-\xi = -\frac{1}{\xi^2}+\frac{\mu}{\left(1-\xi\right)^2}\textrm{sign}\left[1-\xi\right]. \label{xidiffeq}
\end{equation}
Here $\mu \equiv M_{\rm c}/M_{\bullet}$ is the ratio of the mass of the core to the mass of the SMBH.

Integrating over the cross-sectional area of the stream, the continuity equation can be written

\begin{equation}
\frac{\partial m}{\partial t}+v\frac{\partial m}{\partial r} = 0, 
\end{equation}
where $m(r,t)$ is the integrated mass contained within the stream from 0 to $r$. In terms of $\xi$ and $\tau$, this becomes{, after using Equation \eqref{xioftau},}

\begin{equation}
\frac{\partial m}{\partial \tau}+\left(\frac{v}{V}-\xi\right)\frac{\partial m}{\partial \xi} = \frac{\partial m}{\partial \tau}+\frac{\partial \xi}{\partial \tau}\frac{\partial m}{\partial \xi}= \frac{\partial m}{\partial \tau}\bigg{|}_{\xi_0} = 0,
\end{equation}
and hence $m = m(\xi_0)$, where $\xi_0$ is the initial position of a Lagrangian fluid element. The fallback rate onto the black hole can then be found by calculating $\xi_0(\xi,\tau)$ from Equation \eqref{xidiffeq}, adopting a specific form for the mass profile of the stream, and letting $\xi \rightarrow 0$.

\subsection{Asymptotic evolution}
\label{sec:asymptotic}
In general, the solution for the fallback rate must be computed numerically by solving Equation \eqref{xidiffeq}. However, at late times the fallback rate is dominated by gas near the marginally-bound radius, which we will denote by $\xi_{\rm m}(\xi_{0,\rm m},\tau)$, where $\xi_{0, \rm m}$ is the initial position of the marginally-bound fluid element. We can expand the solution for $\xi$ about this radius:

\begin{equation}
\xi(\xi_0,\tau) \simeq \xi_{\rm m}(\tau)+\left(\xi_0-\xi_{\rm 0, m}\right)\xi_{1}(\tau). 
\label{xiexp}
\end{equation}
The marginally-bound radius satisfies $\xi_{\rm m}(\tau \gg 1) \rightarrow {\rm const} \equiv \xi_{\infty}$. Inserting Equation \eqref{xiexp} into Equation \eqref{xidiffeq} and setting $\xi_0 \equiv \xi_{0,\rm m}$, we find that $\xi_{\infty}$ satisfies

\begin{equation}
\xi_{\infty} = \frac{1}{\xi_{\infty}^2}-\frac{\mu}{\left(1-\xi_{\infty}\right)^2}. \label{xiinf}
\end{equation}
Note that if $\mu \equiv 0$ we have $\xi_{\infty} = 1$, i.e., the marginally-bound radius coincides with $r = R$. For non-zero $\mu$, $\xi_{\infty}$ will be slightly less than one because of the competing effects of the gravitational field of the core and that of the SMBH. The first-order terms in $\xi_0-\xi_{0, \rm m}$ give for the bound material ($1-\xi>0$) 

\begin{equation}
2\frac{d^2{\xi}_1}{d\tau^2}+\frac{d{\xi}_1}{d\tau}-\xi_1 = \frac{2}{\xi_{\rm m}^3}\xi_1+\frac{2\mu}{\left(1-\xi_{\rm m}\right)^3}\xi_1.
\end{equation}
When $\xi_{\rm m} \rightarrow \xi_{\infty}$, the solution for $\xi_1$ is 

\begin{equation}
\xi_1 = C_+e^{\omega_+\tau}+C_-e^{\omega_-\tau},
\end{equation}
where $C_{\pm}$ are constants and 

\begin{equation}
\omega_{\pm} = \frac{1}{4}\left(-1\pm\sqrt{9+\frac{16}{\xi_{\infty}^3}+\frac{16\mu}{\left(1-\xi_{\infty}\right)^3}}\right). \label{omegapm}
\end{equation}
Inverting Equation \eqref{xiexp} and taking the limits $\xi \rightarrow 0$ and $\tau \rightarrow \infty$ gives

\begin{equation}
\xi_0 \simeq \xi_{0,\rm m}-\frac{\xi_{\infty}}{C_{+}e^{\omega_{+}\tau}},
\end{equation}
and hence

\begin{equation}
\dot{m} \simeq \frac{V}{R}\frac{\omega_{+}\xi_{\infty}}{C_{+}e^{\omega_+\tau}}\frac{dm}{d\xi_0}\bigg{|}_{\xi_{0 \rm m}} \propto t^{-1-\frac{2}{3}\omega_{+}} \propto t^{n_{\infty}}, \label{mdotasymp}
\end{equation}
where $n_{\infty} \equiv -1-2\omega_+/3$. If $\mu \equiv 0$, which implies $\xi_{\infty} = 1$, Equation \eqref{omegapm} gives $\omega_+ = 1$ and hence $\dot{m} \propto t^{-5/3}$ ($n_{\infty} = -5/3$). In general, we can determine $\xi_{\infty}(\mu)$ by solving Equation \eqref{xiinf}, and Equation \eqref{omegapm} then gives $\omega_+$, and hence $n_{\infty}$, as a function of the core mass. {The exact solution for $\xi_{\infty}(\mu)$ must be calculated numerically, but in the limit of $\mu \ll 1$ -- which is a good approximation for most TDEs where $M_{\rm c}/M_{\bullet} \lesssim 10^{-5}$ -- it can be shown from Equation \eqref{xiinf} that $\xi_{\infty}$ is given by}

\begin{equation}
\xi_{\infty} = 1-\left(\frac{\mu}{3}\right)^{1/3}+\frac{3}{8}\left(\frac{\mu}{3}\right)^{2/3}+\mathcal{O}(\mu),
\end{equation}
{and inserting this expression into Equation \eqref{omegapm} and using the fact that $n_{\infty} = -1-2/3\omega_{+}$ yields}

\begin{equation*}
n_{\infty} = -1-\frac{\sqrt{73}-1}{6}-\frac{17}{2\sqrt{73}}\left(\frac{\mu}{3}\right)^{1/3}+\mathcal{O}(\mu^{2/3}).
\end{equation*}
\begin{equation}
\simeq -2.257-0.690\mu^{1/3}+\mathcal{O}(\mu^{2/3}).  \label{napp}
\end{equation}

\begin{figure*}[htbp] 
   \centering
   \includegraphics[width=0.495\textwidth]{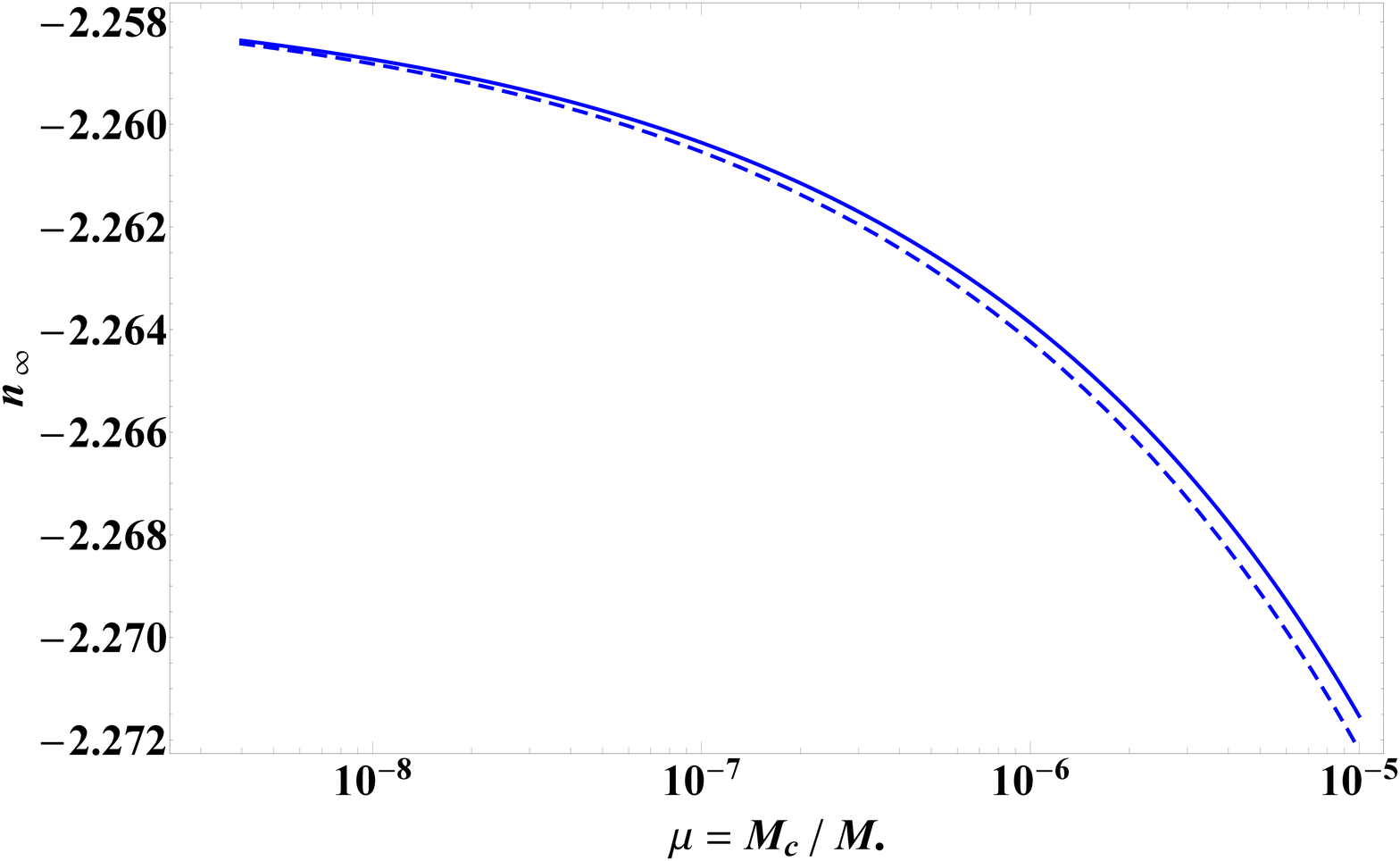} 
   \includegraphics[width=0.495\textwidth]{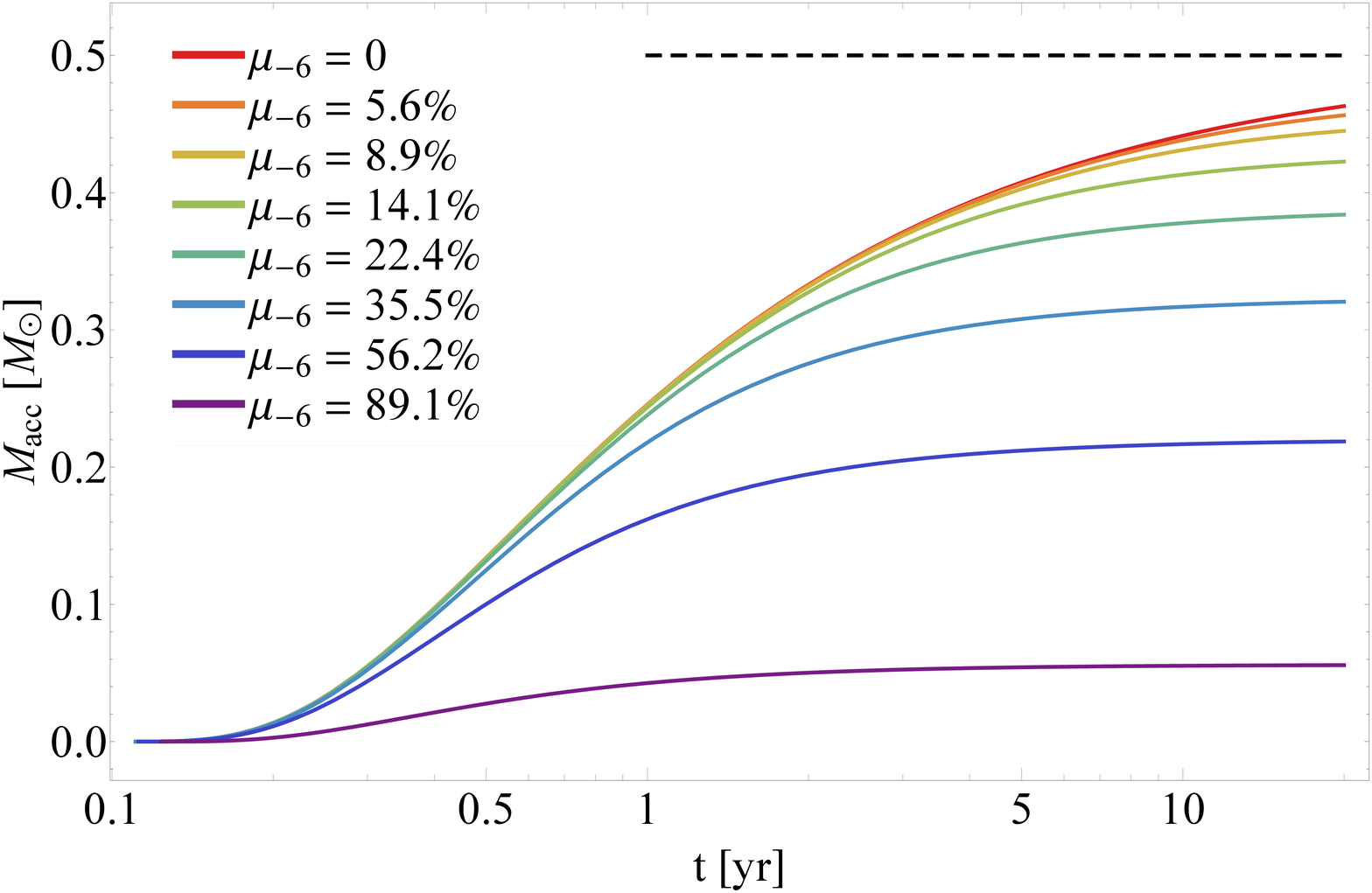} 
   \caption{Left: The asymptotic, temporal power-law index of the fallback rate onto the black hole as a function of the ratio of the core mass to the black hole mass, $\mu$. {The solid line is the numerical solution to Equation \eqref{xiinf}, while the dashed line is the approximation given in Equation \eqref{napp}}. We see that over five orders of magnitude in $\mu$, the power-law index changes by $\sim 0.01$, and hence the asymptotic power-law is effectively independent of the core mass and given by $n_{\infty} \simeq -9/4$. Right: The mass accreted by a $10^6M_{\odot}$ SMBH black hole in Solar masses as a function of time in years, obtained by integrating the differential equation for $\xi(\xi_0,\tau)$ (Equation \ref{xidiffeq}). Here the initial mass profile was approximated from the frozen-in approximation for a Solar-like star modeled as a $\gamma = 5/3$ polytrope. Each curve corresponds to the value of $\mu$ shown in the legend, where $\mu_{-6} = \mu/10^{-6}$ (which is the fraction of the stellar mass that is contained in the surviving core), and the black, dashed curve shows $M_{\rm acc} = 0.5M_{\odot}$ for reference. At late times the total accreted mass is $M_{\rm acc}(t\rightarrow\infty) = (1-\mu_{-6})/2$.  }
   \label{fig:n_infinity}
\end{figure*}

The left panel of Figure \ref{fig:n_infinity} shows the function $n_{\infty}(\mu)$ {obtained numerically from the solution to Equation \eqref{xiinf} (solid) and the analytic approximation given by Equation \eqref{napp} (dashed)}. We see that over a wide range of $\mu$, the value of $n_{\infty}$ varies from $-2.258 \lesssim n_{\infty} \lesssim -2.271$. The asymptotic fallback rate is therefore \emph{effectively independent} of the mass of the core that survives the TDE, and can be written in the particularly simple rational form as\footnote{From Figure \ref{fig:n_infinity}, a slightly better approximation is clearly $n_{\infty} = -2.26$, but this has the aesthetically-unappealing and cumbersome rational form of $n_{\infty} = -113/50$.} $n_{\infty} \simeq -9/4 = -2.25$. 

\begin{figure*}[htbp] 
   \centering
   \includegraphics[width=0.495\textwidth]{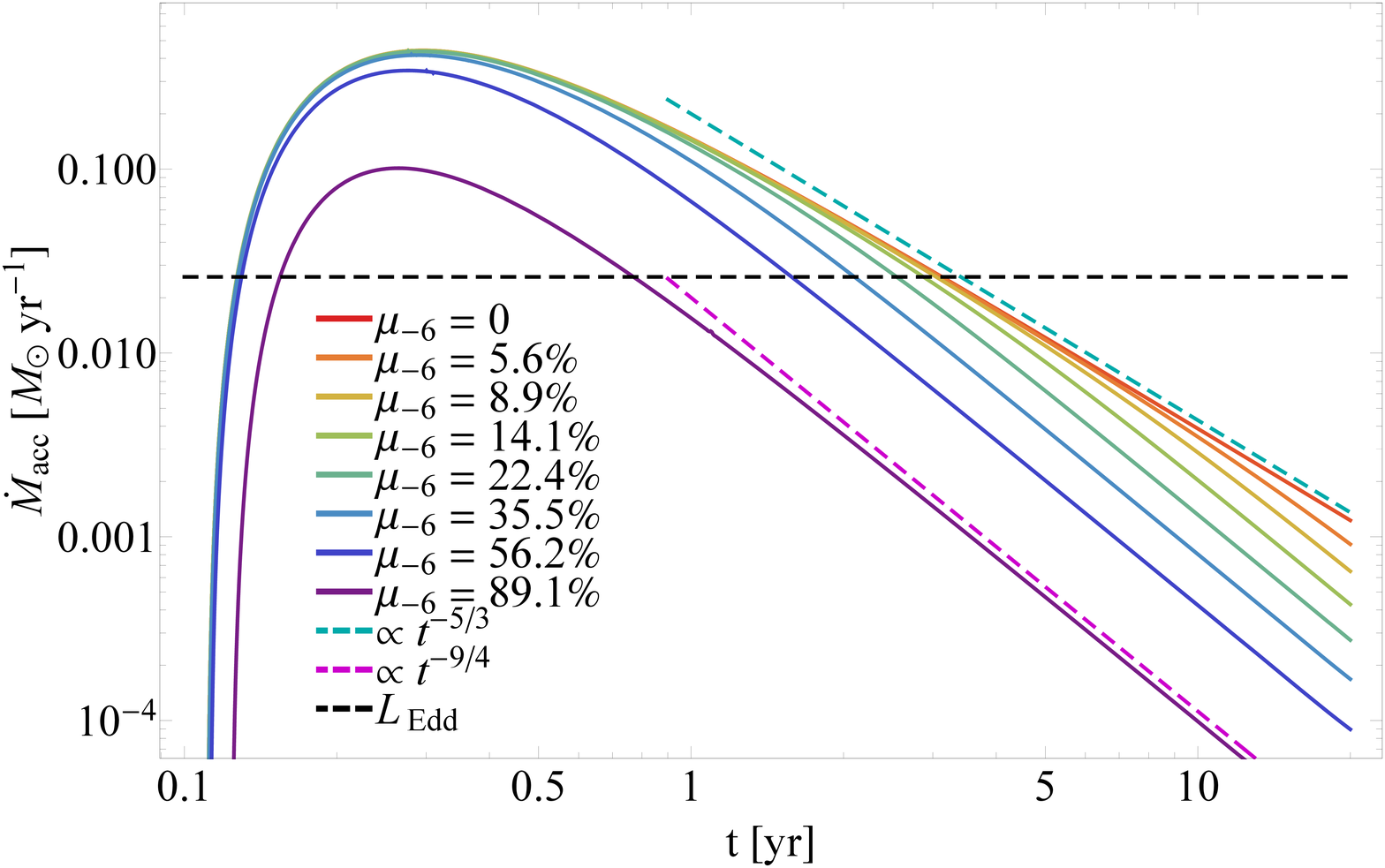} 
   \includegraphics[width=0.495\textwidth]{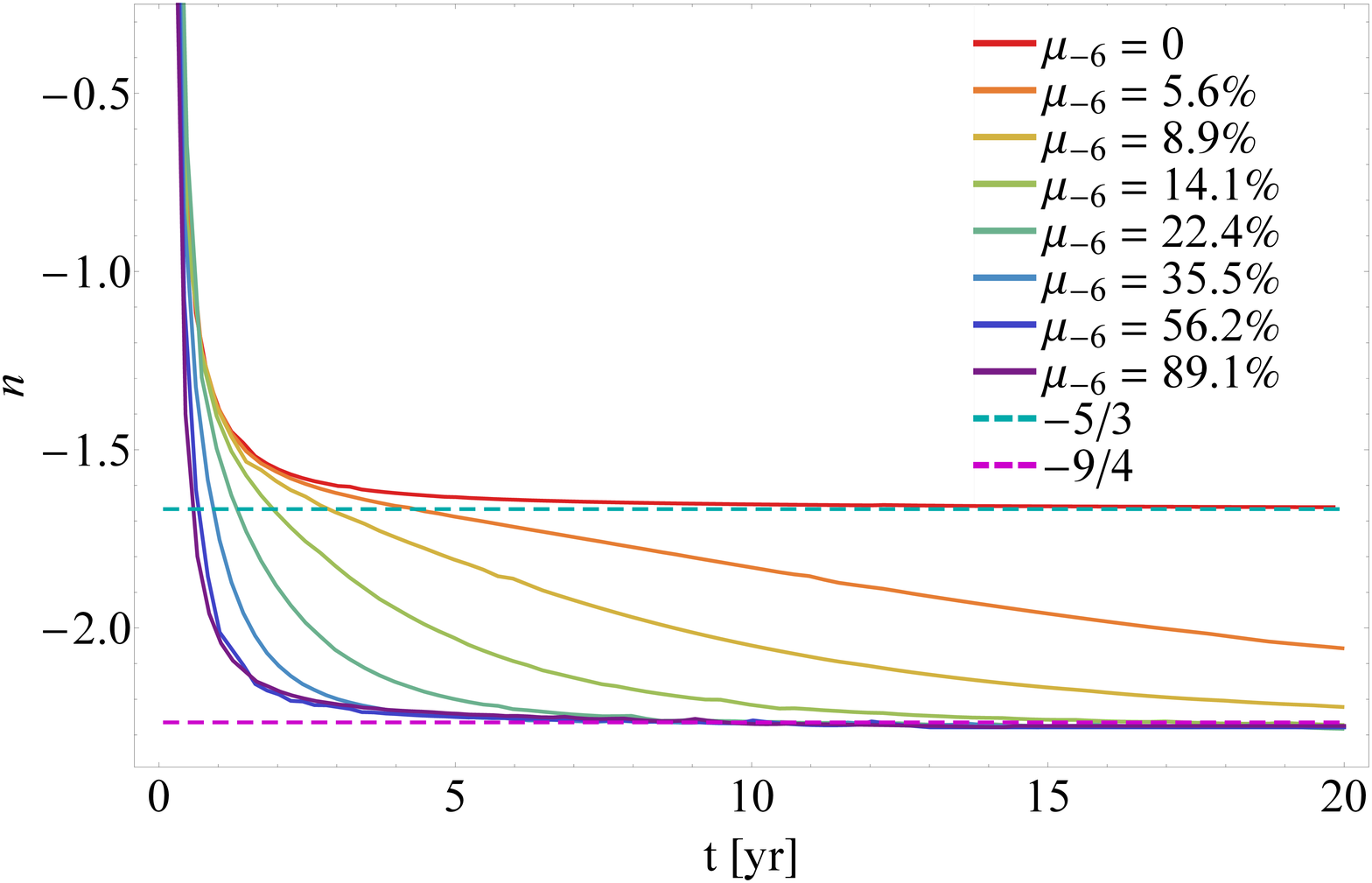} 
   \caption{Left: The fallback rate in Solar masses per year onto a $10^6M_{\odot}$ SMBH as a function of time in years, computed from the temporal derivative of the accreted mass (the right panel of Figure \ref{fig:n_infinity}), where the initial mass profile was approximated from the frozen-in approximation for a Solar-like star modeled as a $\gamma = 5/3$ polytrope. Different curves are appropriate to the values of $\mu_{-6} = \mu/10^{-6}$ (which is the percentage of the initial stellar mass contained in the core) shown in the legend, and the dashed lines give the asymptotic fallback rates for a stream with (magenta, $\propto t^{-9/4}$) and without (cyan, $\propto t^{-5/3}$) a surviving core. The horizontal, dashed line gives the Eddington luminosity of the black hole assuming a radiative efficiency of 10\% (i.e., the accretion luminosity is $L_{\rm acc} = 0.1 \dot{M}_{\rm acc}$) and an electron scattering opacity of 0.34 cm$^{2}$ g$^{-1}$. Right: The instantaneous temporal power-law index of the fallback rate, $n = d\ln\dot{M}/d\ln t$, for the values of $\mu$ shown in the legend. The dashed cyan and magenta lines show the values $-5/3$ and $-9/4$, which give the asymptotic scaling of $n$ without and with a surviving core, respectively. }
   \label{fig:mdots}
\end{figure*}

\subsection{Impulse Approximation}
\label{sec:impulse}
The right panel of Figure \ref{fig:n_infinity} illustrates the total mass accreted by a $10^6M_{\odot}$ SMBH computed by solving Equation \eqref{xidiffeq} for the range of $\mu_{-6} \equiv \mu/10^{-6}$ shown in the legend. Here we assumed that when $R = r_{\rm t}$ the particle velocities were equal to that of the center of mass and the stellar structure was given by a $\gamma = 5/3$ polytrope of radius $R_{\star} = R_{\odot}$ and mass $M_{\star} = M_{\odot}$, in which case \citep{lodato09, golightly19}

\begin{equation}
\frac{dm}{d\xi_0} = 2\pi R_{\star}^3\int_{|1-q^{1/3}\xi_0|}^{1}\rho\eta d\eta. \label{dmdxi0}
\end{equation}
Here $\eta = R/R_{\star}$ is the normalized spherical radius measured from the stellar center of mass, $q = M_{\bullet}/M_{\star} = 10^6$, and $\rho$ is the density of the polytrope. 
In addition, to ensure that the mass of the core is equal to the mass of the original polytrope that is within the radii bound to the core, we allowed the core to ``form'' at some later time $\tau_{\rm c}$ (i.e., we let $\mu \rightarrow \mu\times \Theta[\tau-\tau_{\rm c}]$ in Equation \ref{xidiffeq} with $\Theta$ the Heaviside step function). The value of $\tau_{\rm c}$ was found by requiring that the total mass accreted by the core was equal to the initial core mass.

This approximation clearly cannot be valid at early times in partial TDEs, as the stellar center of mass never actually reaches $r_{\rm t}$ in such encounters (at least for $\gamma = 5/3$ polytropes, for which the critical $\beta$ for full disruption is $\beta \simeq 0.9$; \citealt{guillochon13, mainetti17}), the velocity is predominantly azimuthal (not radial), and some fraction of the stellar mass remains in the core at all times. However, in a realistic encounter, one can imagine that such a frozen-in model is approximately valid when the center of mass is at a radius somewhat larger than $r_{\rm t}$, the density distribution is slightly elongated from the initial stellar one, and the amount of mass in the core has saturated to a roughly constant value. Therefore, by starting with a spherical arrangement of mass at an artificially small radius and forcing the core to form at a later time, we allow the system to conform to a distribution that is similar to a more realistic one at a larger radius.

More importantly, this approximation gives a reasonable distribution of $\xi_0$ and $\dot{\xi}(\tau=0) = 1-\xi_0$ (see Equation \ref{xidef}) that can be used to integrate Equation \eqref{xidiffeq} and assess the temporal variation of the fallback. The left panel of Figure \ref{fig:mdots} shows the accretion rate $\dot{M}_{\rm acc}$ in Solar masses per year as a function of time in years, where each curve is obtained by taking the time derivative of the respective curve in the right panel of Figure \ref{fig:n_infinity}. We see that the fallback rates rise to peak on a timescale of months, and the curve with $\mu \equiv 0$ asymptotically follows $\dot{m} \propto t^{-5/3}$ (cyan, dashed curve). For partial TDEs with a core mass $\mu_{-6} \lesssim 15\%$, each fallback rate closely tracks the core-less curve for a considerable amount of time, but nonetheless eventually exhibits deviations from and falls below the $\mu_{-6} = 0$ curve. For partial TDEs with a more sizable core mass, the fallback curves never resemble the core-less one and follow the $\propto t^{-9/4}$ asymptotic scaling (magenta, dashed curve) at a time of roughly one year. In every case the fallback rate initially exceeds the Eddington luminosity of the SMBH (black, dashed line, assuming a radiative efficiency of 10\%). 

The right panel of Figure \ref{fig:mdots} shows the logarithmic derivative of the fallback rate, $n(t) = d\ln \dot{M}/d\ln t$, which tracks the instantaneous temporal power-law index of the fallback curves in the left panel of this figure. As expected, the power-law index of the core-less TDE asymptotes to $n = -5/3$ at late times, and reaches a value of $n = -1.5$ by a time of roughly one year. Solutions with a substantial core mass ($\gtrsim 15\%$ of the initial stellar mass) quickly asymptote to $n = -9/4$ and reach $n = -2$ by approximately one year. When the core mass is on the order of 10\% of the initial stellar mass, the solutions spend a considerable amount of time being matched by neither $n = -5/3$ nor $n = -9/4$, though for $\mu_{-6} = 5.6\%$ (and, to a lesser extent, $\mu_{-6} = 8.9\%$) the solution is well-matched by $n = -5/3$ for the first $\sim 5$ years (as is evident from the right panel of this figure). 

\section{Summary and Conclusions}
\label{sec:summary}
In this paper we developed a model to determine the fallback rate from a partial tidal disruption event, which -- in addition to producing a stream of debris that feeds the black hole -- leaves a bound core that survives the tidal encounter. Using this model, we calculated the asymptotic rate at which material returns to the black hole as a function of the mass of the surviving core; we found that this rate scales with time $t$ as $t^{n_{\infty}}$ with $n_{\infty} = -2.26\pm 0.01$, with the upper and lower limits on $n_{\infty}$ encompassing orders of magnitude in the remnant core mass (see Figure \ref{fig:n_infinity}). This model therefore predicts that the fallback rate from partial TDEs effectively approaches a universal power-law scaling of $\dot{M} \propto t^{-9/4}$. 

We also employed a variation of the impulsive, or frozen-in model as described in \citet{lodato09} to assess the rate at which the fallback curves from partial TDEs approach this universal scaling. We found that, when the core contains $\lesssim 15\%$ of the mass of the initial star, the fallback rate can take greater than $\sim 3-5$ years before any appreciable deviation from the fallback rate from a full disruption arises (see Figure \ref{fig:mdots}). However, for disruptions that leave a more substantial core, the fallback rate exhibits noticeable differences at much earlier times and the rate quickly asymptotes to $\propto t^{-9/4}$. 

Our results suggest that TDEs produce a dichotomy of fallback rates, with partial (full) disruptions generating an asymptotic accretion rate that scales as $t^{-9/4}$ ($t^{-5/3})$. If tidally-disrupted stars are uniformly scattered in angular momentum space (i.e., come from a full loss cone; \citealt{cohn78, stone16}), then the distribution function of the pericenter distance of tidally-disrupted stars that produce flares is $f_{\beta} = \beta_{\rm min}/\beta^2$, where $\beta_{\rm min}$ is the mimimum $\beta$ at which some mass is tidally stripped from the star. The fraction of partial disruptions is then $N_{\rm partial} = 1-\beta_{\rm min}/\beta_{\rm crit}$, where $\beta_{\rm crit}$ separates full and partial disruptions. \citet{mainetti17} find that $\beta_{\rm min} = 0.5$ and $\beta_{\rm crit} = 0.95$ for $\gamma = 5/3$ polytropes, while $\beta_{\rm min} =0.6$ and $\beta_{\rm crit} = 2$ for a $\gamma = 4/3$ polytrope. We therefore find that $N_{\rm partial} \simeq 47\%$ of low-mass stars with $\gamma = 5/3$ result in partial disruptions, while $N_{\rm partial} \simeq 70\%$ of high mass stars (or more highly evolved stars that have developed denser cores) generate partial TDEs and therefore have fallback rates that scale as $t^{-9/4}$. While the total mass accreted in partial TDEs is less and self-intersection is more difficult owing to the reduced general-relativistic precession, both of which conspire to yield a less-luminous flare as compared to a full disruption\footnote{Though we note that the accretion rates in Figure \ref{fig:mdots} are super-Eddington for $\sim 1$ yr even for the most massive core, and the large $\beta$'s required to fully disrupt a denser, $\gamma = 4/3$ polytrope yield a large periapsis advance angle that would facilitate self-intersection.}, we expect that a substantial fraction of TDEs detected by future facilities, such as the Large Synoptic Survey Telescope \citep{ivezic19}, will produce flares that decay as $t^{-9/4}$. 

Additionally, there are physical systems that can masquerade as TDEs by producing a bolometric luminosity that scales as $t^{-5/3}$, one example being the fallback onto a black hole or neutron star following a core-collapse supernova \citep{chevalier89, dexter13, fernandez18}. However, since the density of the ejected hydrogen envelope is orders of magnitude below that of a tidally-disrupted debris stream and declines much more rapidly with time\footnote{Under the frozen-in approximation, the density of the ejected shell from a supernova declines as $t^{-8/3}$, as compared to the $t^{-2}$ decline for a gas-pressure-dominated debris stream \citep{coughlin16b}.}, it is much more difficult to form a recollapsed ``shell'' of material out of the supernova ejecta that could then modify the Keplerian energy distribution of the infalling debris. Therefore, the decline of an observed lightcurve as $t^{-9/4}$ could be seen as a more robust indicator of a TDE than the canonical rate of $t^{-5/3}$.

{We also emphasize an important, physical point regarding the nature of our model and our results in comparison to others in the literature: previous calculations of TDE fallback rates have 1) {presupposed the existence of a conserved, Lagrangian energy for the gas parcels within the debris stream;} 2) {assumed that the conserved energy is given by that of a Keplerian orbit}; and 3) used the relationship between the energy of a Keplerian orbit and {the orbital timescale}, which results in a fallback rate of the form (e.g., Equation 1 of \citealt{guillochon13} and Equation 7 of \citealt{goicovic19})}

\begin{equation}
\dot{m} \propto t^{-5/3}\frac{dM}{dE}. \label{mdotwrong}
\end{equation}
{Crucially, {the existence of a conserved energy -- upon which this expression is predicated -- is only guaranteed when the gravitational potential is time-independent. However, the time independence of the potential is manifestly violated when a bound core exerts a gravitational field at the (temporally-evolving) marginally-bound radius, and the lack of a conserved, Lagrangian energy is evident from Equation \eqref{eq1}.}

{Because of the non-existence of a conserved energy, Equation \eqref{mdotwrong} does not lead to a self-consistent determination of the asymptotic fallback rate. For example, if one exploits the one-to-one mapping between the Lagrangian position within the stream and the energy, which we reiterate \emph{is only valid when the energy of each orbit is absolutely conserved}, then at asymptotically late times the function $dM/dE$ in Equation \eqref{mdotwrong} becomes}

\begin{equation}
\frac{dM}{dE} \rightarrow \frac{dM}{dE}\bigg{|}_{E = 0} \propto \frac{dM}{d\xi_0}\bigg{|}_{\xi_{\rm 0,m}},
\end{equation}
where $\xi_0$ is the initial location within the stream and $\xi_{\rm 0,m}$ is the marginally-bound radius. This result in conjunction with Equation \eqref{mdotwrong} suggests that the fallback rate should eventually approach $t^{-5/3}${, independent of the presence of a bound core. On the other hand, one can simply use the Keplerian energy $E = v^2/2-GM_{\bullet}/r$ to calculate the energy distribution within the stream and use the relationship $E \propto t^{-2/3}$. However, the presence of the bound core modifies the location of the marginally-bound radius to the black hole within the stream, and the Keplerian energy of that radius is slightly negative. Therefore, if one calculates the fallback rate of {only} the material that is bound to the black hole and excludes mass at larger radii, then one finds that the fallback abruptly terminates at a finite time (and that time depends on when the Keplerian energy distribution is measured post-disruption).} 

{The origin of these contradictory conclusions is the \emph{assumption} of the existence of a {conserved,} Keplerian energy for the gas parcels within the debris stream. This assumption is likely to be approximately valid for the early-time fallback, for which the tidally-disrupted debris has not been greatly affected by the presence of the core (though this statement must also depend on the mass of the core). However, this assumption cannot be valid for the late-time fallback curve, as material that falls back at ever-later times is increasingly modified by the gravitational potential of the core. In contrast, our analysis only exploits the fact that the integrated mass within the stream is only a function of Lagrangian position, which is tantamount to the statement that particle orbits cannot cross, and is therefore upheld even when there is no conserved Lagrangian energy to exploit.} 

{The fallback rate onto the black hole is predicted to closely track the accretion rate for a timescale of tens of years, after which time the accretion rate should be viscously delayed \citep{ulmer99} and follows a shallower power-law decline of $t^{-1.2}$ \citep{cannizzo90,shen14}; this transition is seen in the late-time observations of TDEs, as found by \citet{vanvelzen19}, and the analysis of \citet{auchettl17} suggests that viscous delays in X-ray bright TDEs may occur even earlier. \citet{lodato11} also argued that the early-time X-ray emission should track the fallback rate, and the analysis of 14 observed TDEs by \citet{mockler19} found that the viscous delay in that subset was extremely small in the UV and optical (implying that the fallback rate tracked the emission in those bands).} {It is therefore reasonable to expect the lightcurves of at least some fraction of partial TDEs to exhibit the $\propto t^{-9/4}$ decay predicted here.}

{Our simple model employed to assess the rate at which partial TDEs approach a $t^{-9/4}$ scaling (Figure \ref{fig:mdots}) suggests that this power-law is reached (to within a small factor) on a timescale of months to years when the partial disruption leaves a substantial fraction of mass ($\gtrsim 90\%$ of the initial stellar mass) contained in the core, while the transition occurs closer to 5-10 years post-disruption when the core retains $\lesssim 10\%$ of the initial mass of the star. In general, the rapidity with which the fallback rate assumes this power-law depends on the distribution of mass along the stream $dm/d\xi_0$, and the closer this distribution becomes to uniform, the earlier that this power-law dependence is reached (see Equation \ref{mdotasymp}). In our model, the original, polytropic density structure of the star is maintained by the stream, and it is this additional variation in $dm/d\xi_0$ that provides the delay in approaching the $t^{-9/4}$ decline. However, it has been found \citep{lodato09, guillochon13} that the distribution of $dm/d\xi_0$ (equivalent to $dm/dE$ when the energy is a conserved Lagrangian variable) becomes considerably flatter when the calculation is done numerically, and \citet{coughlin16a} argued that this flattening is due to the effects of self-gravity at early times. We therefore expect the $t^{-9/4}$ decline to be reached \emph{earlier} than predicted in Figure \ref{fig:mdots} when detailed simulations are performed, and the results of \citet{golightly19b} -- who directly measured the rate of return of bound debris to the black hole and did not rely on an energy-period relation to forward-predict the fallback rate -- substantiate this claim (see their Figures 4 and 6, which demonstrate that the $t^{-9/4}$ decline can be reached almost immediately after peak).}

Our model only accounts for the gravitational influence of the core and the SMBH, and ignores the pressure and self-gravity of the stream itself, the fact that the position of the core may not follow a zero-energy orbit exactly \citep{fabian75,rasio91,manukian13,sacchi19}, and higher-order moments of the gravitational field of the core{; the latter feature must, for very small values of the core mass, become important as the marginally-bound radius approaches the surface of the core itself. When the surface of the core and the marginally-bound radius start to coincide, the mass distribution -- as concerns the marginally-bound fluid element -- is better approximated by a continuum and not that of a point mass. In fact, the divergence of the gravitational field of the point mass as $\xi_{\infty} \rightarrow 1$ is why $n_{\infty} \neq -5/3$ as $\mu \rightarrow 0$ in the left panel of Figure \ref{fig:n_infinity} {(i.e., a point mass will always maintain a finite Hill sphere from which the returning debris leaves, which modifies the temporal power-law index of the late-time fallback rate)}. We also assumed a very simple distribution of mass to calculate the temporal dependence of the fallback from a partial TDE. We plan to analyze the influence of these additional, physical effects on the late-time (and early-time) fallback from partial TDEs in a future investigation. 

\acknowledgements{We thank Patrick Miles for useful discussions that added to the quality of this paper. ERC acknowledges support from the Lyman Spitzer Jr.~Postdoctoral Fellowship, and from NASA through the Einstein Fellowship Program, grant PF6-170170, and the Hubble Fellowship, grant \#HST-HF2-51433.001-A awarded by the Space Telescope Science Institute, which is operated by the Association of Universities for Research in Astronomy, Incorporated, under NASA contract NAS5-26555. CJN is supported by the Science and Technology Facilities Council (grant number ST/M005917/1).}


\begin{thebibliography}{}
\expandafter\ifx\csname natexlab\endcsname\relax\def\natexlab#1{#1}\fi
\providecommand{\url}[1]{\href{#1}{#1}}
\providecommand{\dodoi}[1]{doi:~\href{http://doi.org/#1}{\nolinkurl{#1}}}
\providecommand{\doeprint}[1]{\href{http://ascl.net/#1}{\nolinkurl{http://ascl.net/#1}}}
\providecommand{\doarXiv}[1]{\href{https://arxiv.org/abs/#1}{\nolinkurl{https://arxiv.org/abs/#1}}}

\bibitem[{{Arcavi} {et~al.}(2014){Arcavi}, {Gal-Yam}, {Sullivan}, {Pan},
  {Cenko}, {Horesh}, {Ofek}, {De Cia}, {Yan}, \& {Yang}}]{arcavi14}
{Arcavi}, I., {Gal-Yam}, A., {Sullivan}, M., {et~al.} 2014, \apj, 793, 38,
  \dodoi{10.1088/0004-637X/793/1/38}

\bibitem[{{Auchettl} {et~al.}(2017){Auchettl}, {Guillochon}, \&
  {Ramirez-Ruiz}}]{auchettl17}
{Auchettl}, K., {Guillochon}, J., \& {Ramirez-Ruiz}, E. 2017, \apj, 838, 149,
  \dodoi{10.3847/1538-4357/aa633b}

\bibitem[{{Blagorodnova} {et~al.}(2017){Blagorodnova}, {Gezari}, {Hung},
  {Kulkarni}, {Cenko}, {Pasham}, {Yan}, {Arcavi}, {Ben-Ami}, \&
  {Bue}}]{blagorodnova17}
{Blagorodnova}, N., {Gezari}, S., {Hung}, T., {et~al.} 2017, \apj, 844, 46,
  \dodoi{10.3847/1538-4357/aa7579}

\bibitem[{{Bloom} {et~al.}(2011){Bloom}, {Giannios}, {Metzger}, {Cenko},
  {Perley}, {Butler}, {Tanvir}, {Levan}, {O'Brien}, \& {Strubbe}}]{bloom11}
{Bloom}, J.~S., {Giannios}, D., {Metzger}, B.~D., {et~al.} 2011, Science, 333,
  203, \dodoi{10.1126/science.1207150}

\bibitem[{{Bonnerot} {et~al.}(2016){Bonnerot}, {Rossi}, {Lodato}, \&
  {Price}}]{bonnerot16}
{Bonnerot}, C., {Rossi}, E.~M., {Lodato}, G., \& {Price}, D.~J. 2016, \mnras,
  455, 2253, \dodoi{10.1093/mnras/stv2411}

\bibitem[{{Cannizzo} {et~al.}(1990){Cannizzo}, {Lee}, \&
  {Goodman}}]{cannizzo90}
{Cannizzo}, J.~K., {Lee}, H.~M., \& {Goodman}, J. 1990, \apj, 351, 38,
  \dodoi{10.1086/168442}

\bibitem[{{Cenko} {et~al.}(2012){Cenko}, {Krimm}, {Horesh}, {Rau}, {Frail},
  {Kennea}, {Levan}, {Holland}, {Butler}, \& {Quimby}}]{cenko12}
{Cenko}, S.~B., {Krimm}, H.~A., {Horesh}, A., {et~al.} 2012, \apj, 753, 77,
  \dodoi{10.1088/0004-637X/753/1/77}

\bibitem[{{Chevalier}(1989)}]{chevalier89}
{Chevalier}, R.~A. 1989, \apj, 346, 847, \dodoi{10.1086/168066}

\bibitem[{{Chornock} {et~al.}(2014){Chornock}, {Berger}, {Gezari}, {Zauderer},
  {Rest}, {Chomiuk}, {Kamble}, {Soderberg}, {Czekala}, \&
  {Dittmann}}]{chornock14}
{Chornock}, R., {Berger}, E., {Gezari}, S., {et~al.} 2014, \apj, 780, 44,
  \dodoi{10.1088/0004-637X/780/1/44}

\bibitem[{{Cohn} \& {Kulsrud}(1978)}]{cohn78}
{Cohn}, H., \& {Kulsrud}, R.~M. 1978, \apj, 226, 1087, \dodoi{10.1086/156685}

\bibitem[{{Coughlin} \& {Nixon}(2015)}]{coughlin15}
{Coughlin}, E.~R., \& {Nixon}, C. 2015, \apj, 808, L11,
  \dodoi{10.1088/2041-8205/808/1/L11}

\bibitem[{{Coughlin} {et~al.}(2016{\natexlab{a}}){Coughlin}, {Nixon},
  {Begelman}, \& {Armitage}}]{coughlin16b}
{Coughlin}, E.~R., {Nixon}, C., {Begelman}, M.~C., \& {Armitage}, P.~J.
  2016{\natexlab{a}}, \mnras, 459, 3089, \dodoi{10.1093/mnras/stw770}

\bibitem[{{Coughlin} {et~al.}(2016{\natexlab{b}}){Coughlin}, {Nixon},
  {Begelman}, {Armitage}, \& {Price}}]{coughlin16a}
{Coughlin}, E.~R., {Nixon}, C., {Begelman}, M.~C., {Armitage}, P.~J., \&
  {Price}, D.~J. 2016{\natexlab{b}}, \mnras, 455, 3612,
  \dodoi{10.1093/mnras/stv2511}

\bibitem[{{Dexter} \& {Kasen}(2013)}]{dexter13}
{Dexter}, J., \& {Kasen}, D. 2013, \apj, 772, 30,
  \dodoi{10.1088/0004-637X/772/1/30}

\bibitem[{{Evans} \& {Kochanek}(1989)}]{evans89}
{Evans}, C.~R., \& {Kochanek}, C.~S. 1989, \apj, 346, L13,
  \dodoi{10.1086/185567}

\bibitem[{{Fabian} {et~al.}(1975){Fabian}, {Pringle}, \& {Rees}}]{fabian75}
{Fabian}, A.~C., {Pringle}, J.~E., \& {Rees}, M.~J. 1975, \mnras, 172, 15,
  \dodoi{10.1093/mnras/172.1.15P}

\bibitem[{{Fern{\'a}ndez} {et~al.}(2018){Fern{\'a}ndez}, {Quataert},
  {Kashiyama}, \& {Coughlin}}]{fernandez18}
{Fern{\'a}ndez}, R., {Quataert}, E., {Kashiyama}, K., \& {Coughlin}, E.~R.
  2018, \mnras, 476, 2366, \dodoi{10.1093/mnras/sty306}

\bibitem[{{Gallegos-Garcia} {et~al.}(2018){Gallegos-Garcia}, {Law-Smith}, \&
  {Ramirez-Ruiz}}]{gallegos-garcia18}
{Gallegos-Garcia}, M., {Law-Smith}, J., \& {Ramirez-Ruiz}, E. 2018, \apj, 857,
  109, \dodoi{10.3847/1538-4357/aab5b8}

\bibitem[{{Gezari} {et~al.}(2012){Gezari}, {Chornock}, {Rest}, {Huber},
  {Forster}, {Berger}, {Challis}, {Neill}, {Martin}, \& {Heckman}}]{gezari12}
{Gezari}, S., {Chornock}, R., {Rest}, A., {et~al.} 2012, \nat, 485, 217,
  \dodoi{10.1038/nature10990}

\bibitem[{{Goicovic} {et~al.}(2019){Goicovic}, {Springel}, {Ohlmann}, \&
  {Pakmor}}]{goicovic19}
{Goicovic}, F.~G., {Springel}, V., {Ohlmann}, S.~T., \& {Pakmor}, R. 2019,
  \mnras, 487, 981, \dodoi{10.1093/mnras/stz1368}

\bibitem[{{Golightly} {et~al.}(2019{\natexlab{a}}){Golightly}, {Coughlin}, \&
  {Nixon}}]{golightly19}
{Golightly}, E. C.~A., {Coughlin}, E.~R., \& {Nixon}, C.~J. 2019{\natexlab{a}},
  \apj, 872, 163, \dodoi{10.3847/1538-4357/aafd2f}

\bibitem[{{Golightly} {et~al.}(2019{\natexlab{b}}){Golightly}, {Nixon}, \&
  {Coughlin}}]{golightly19b}
{Golightly}, E. C.~A., {Nixon}, C.~J., \& {Coughlin}, E.~R. 2019{\natexlab{b}},
  arXiv e-prints, arXiv:1907.05895.
\newblock \doarXiv{1907.05895}

\bibitem[{{Guillochon} {et~al.}(2014){Guillochon}, {Manukian}, \&
  {Ramirez-Ruiz}}]{guillochon14}
{Guillochon}, J., {Manukian}, H., \& {Ramirez-Ruiz}, E. 2014, \apj, 783, 23,
  \dodoi{10.1088/0004-637X/783/1/23}

\bibitem[{{Guillochon} \& {Ramirez-Ruiz}(2013)}]{guillochon13}
{Guillochon}, J., \& {Ramirez-Ruiz}, E. 2013, \apj, 767, 25,
  \dodoi{10.1088/0004-637X/767/1/25}

\bibitem[{{Hayasaki} {et~al.}(2013){Hayasaki}, {Stone}, \& {Loeb}}]{hayasaki13}
{Hayasaki}, K., {Stone}, N., \& {Loeb}, A. 2013, \mnras, 434, 909,
  \dodoi{10.1093/mnras/stt871}

\bibitem[{{Hayasaki} {et~al.}(2016){Hayasaki}, {Stone}, \& {Loeb}}]{hayasaki16}
---. 2016, \mnras, 461, 3760, \dodoi{10.1093/mnras/stw1387}

\bibitem[{{Hills}(1975)}]{hills75}
{Hills}, J.~G. 1975, \nat, 254, 295, \dodoi{10.1038/254295a0}

\bibitem[{{Hung} {et~al.}(2017){Hung}, {Gezari}, {Blagorodnova}, {Roth},
  {Cenko}, {Kulkarni}, {Horesh}, {Arcavi}, {McCully}, \& {Yan}}]{hung17}
{Hung}, T., {Gezari}, S., {Blagorodnova}, N., {et~al.} 2017, \apj, 842, 29,
  \dodoi{10.3847/1538-4357/aa7337}

\bibitem[{{Ivezi{\'c}} {et~al.}(2019){Ivezi{\'c}}, {Kahn}, {Tyson}, {Abel},
  {Acosta}, {Allsman}, {Alonso}, {AlSayyad}, {Anderson}, \&
  {Andrew}}]{ivezic19}
{Ivezi{\'c}}, {\v{Z}}., {Kahn}, S.~M., {Tyson}, J.~A., {et~al.} 2019, \apj,
  873, 111, \dodoi{10.3847/1538-4357/ab042c}

\bibitem[{{Kochanek}(1994)}]{kochanek94}
{Kochanek}, C.~S. 1994, \apj, 422, 508, \dodoi{10.1086/173745}

\bibitem[{{Komossa}(2015)}]{komossa15}
{Komossa}, S. 2015, Journal of High Energy Astrophysics, 7, 148,
  \dodoi{10.1016/j.jheap.2015.04.006}

\bibitem[{{Lodato} {et~al.}(2009){Lodato}, {King}, \& {Pringle}}]{lodato09}
{Lodato}, G., {King}, A.~R., \& {Pringle}, J.~E. 2009, \mnras, 392, 332,
  \dodoi{10.1111/j.1365-2966.2008.14049.x}

\bibitem[{{Lodato} \& {Rossi}(2011)}]{lodato11}
{Lodato}, G., \& {Rossi}, E.~M. 2011, \mnras, 410, 359,
  \dodoi{10.1111/j.1365-2966.2010.17448.x}

\bibitem[{{Mainetti} {et~al.}(2017){Mainetti}, {Lupi}, {Campana}, {Colpi},
  {Coughlin}, {Guillochon}, \& {Ramirez-Ruiz}}]{mainetti17}
{Mainetti}, D., {Lupi}, A., {Campana}, S., {et~al.} 2017, \aap, 600, A124,
  \dodoi{10.1051/0004-6361/201630092}

\bibitem[{{Manukian} {et~al.}(2013){Manukian}, {Guillochon}, {Ramirez-Ruiz}, \&
  {O'Leary}}]{manukian13}
{Manukian}, H., {Guillochon}, J., {Ramirez-Ruiz}, E., \& {O'Leary}, R.~M. 2013,
  \apj, 771, L28, \dodoi{10.1088/2041-8205/771/2/L28}

\bibitem[{{Mockler} {et~al.}(2019){Mockler}, {Guillochon}, \&
  {Ramirez-Ruiz}}]{mockler19}
{Mockler}, B., {Guillochon}, J., \& {Ramirez-Ruiz}, E. 2019, \apj, 872, 151,
  \dodoi{10.3847/1538-4357/ab010f}

\bibitem[{{Phinney}(1989)}]{phinney89}
{Phinney}, E.~S. 1989, in IAU Symposium, Vol. 136, The Center of the Galaxy,
  ed. M.~{Morris}, 543

\bibitem[{{Rasio} \& {Shapiro}(1991)}]{rasio91}
{Rasio}, F.~A., \& {Shapiro}, S.~L. 1991, \apj, 377, 559,
  \dodoi{10.1086/170385}

\bibitem[{{Rees}(1988)}]{rees88}
{Rees}, M.~J. 1988, \nat, 333, 523, \dodoi{10.1038/333523a0}

\bibitem[{{Sacchi} \& {Lodato}(2019)}]{sacchi19}
{Sacchi}, A., \& {Lodato}, G. 2019, \mnras, 486, 1833,
  \dodoi{10.1093/mnras/stz981}

\bibitem[{{S{{a}}dowski} {et~al.}(2016){S{{a}}dowski}, {Tejeda}, {Gafton},
  {Rosswog}, \& {Abarca}}]{sadowski16}
{S{{a}}dowski}, A., {Tejeda}, E., {Gafton}, E., {Rosswog}, S., \& {Abarca}, D.
  2016, \mnras, 458, 4250, \dodoi{10.1093/mnras/stw589}

\bibitem[{{Shen} \& {Matzner}(2014)}]{shen14}
{Shen}, R.-F., \& {Matzner}, C.~D. 2014, \apj, 784, 87,
  \dodoi{10.1088/0004-637X/784/2/87}

\bibitem[{{Shiokawa} {et~al.}(2015){Shiokawa}, {Krolik}, {Cheng}, {Piran}, \&
  {Noble}}]{shiokawa15}
{Shiokawa}, H., {Krolik}, J.~H., {Cheng}, R.~M., {Piran}, T., \& {Noble}, S.~C.
  2015, \apj, 804, 85, \dodoi{10.1088/0004-637X/804/2/85}

\bibitem[{{Steinberg} {et~al.}(2019){Steinberg}, {Coughlin}, {Stone}, \&
  {Metzger}}]{steinberg19}
{Steinberg}, E., {Coughlin}, E.~R., {Stone}, N.~C., \& {Metzger}, B.~D. 2019,
  \mnras, 485, L146, \dodoi{10.1093/mnrasl/slz048}

\bibitem[{{Stone} {et~al.}(2013){Stone}, {Sari}, \& {Loeb}}]{stone13}
{Stone}, N., {Sari}, R., \& {Loeb}, A. 2013, \mnras, 435, 1809,
  \dodoi{10.1093/mnras/stt1270}

\bibitem[{{Stone} \& {Metzger}(2016)}]{stone16}
{Stone}, N.~C., \& {Metzger}, B.~D. 2016, \mnras, 455, 859,
  \dodoi{10.1093/mnras/stv2281}

\bibitem[{{Ulmer}(1999)}]{ulmer99}
{Ulmer}, A. 1999, \apj, 514, 180, \dodoi{10.1086/306909}

\bibitem[{{van Velzen} {et~al.}(2019){van Velzen}, {Stone}, {Metzger},
  {Gezari}, {Brown}, \& {Fruchter}}]{vanvelzen19}
{van Velzen}, S., {Stone}, N.~C., {Metzger}, B.~D., {et~al.} 2019, \apj, 878,
  82, \dodoi{10.3847/1538-4357/ab1844}

\end{thebibliography}

\end{document}